\documentclass{pasj00}

\begin{document}
\SetRunningHead{Fujimoto et al. }{Charge-Exchange X-ray emission}
\Received{2006/7/26}
\Accepted{2006/9/9}

\title{Evidence for Solar-Wind Charge-Exchange X-Ray Emission\\
from the Earth's Magnetosheath}

%
\author{%
Ryuichi \textsc{Fujimoto}\altaffilmark{1},
Kazuhisa \textsc{Mitsuda}\altaffilmark{1},
Dan {\sc McCammon}\altaffilmark{2},
Yoh \textsc{Takei}\altaffilmark{1},
Michael \textsc{Bauer}\altaffilmark{3},\\
Yoshitaka \textsc{Ishisaki}\altaffilmark{4},
F. Scott {\sc Porter}\altaffilmark{5},
Hiroya {\sc Yamaguchi}\altaffilmark{6},
Kiyoshi {\sc Hayashida}\altaffilmark{7},
and Noriko Y. {\sc Yamasaki}\altaffilmark{1}\\
}
 \altaffiltext{1}{ISAS/JAXA, 3-1-1 Yoshinodai, Sagamihara 229-8510}
\email{fujimoto@astro.isas.jaxa.jp}
\altaffiltext{2}{University of Wisconsin, Madison, WI 53706, USA}
\altaffiltext{3}{Max-Planck-Institut f\"ur extraterrestrische Physik, D-85740 Garching, Germany}
\altaffiltext{4}{Tokyo Metropolitan University, 1-1 Minami-Osawa, Hachioji, Tokyo 192-0397}
\altaffiltext{5}{NASA Goddard Space Flight Center, Greenbelt, MD 20771, USA}
\altaffiltext{6}{Department of Physics, Faculty of Science, Kyoto University, Sakyo-ku, Kyoto 606-8502}
\altaffiltext{7}{Department of Astrophysics, Faculty of Science, Osaka
University, Toyonaka 560-0043}


\maketitle

\begin{abstract}

We report an apparent detection of the C\emissiontype{VI} $4p$ to $1s$
transition line at 459~eV, during a long-term enhancement (LTE) in the
{\it Suzaku} north ecliptic pole (NEP) observation of 2005 September 2.
The observed intensity of the line is comparable to that of the
C\emissiontype{VI} $2p$ to $1s$ line at 367~eV. This is strong evidence
for the charge-exchange process. In addition to the C\emissiontype{VI}
lines, emission lines from O\emissiontype{VII}, O\emissiontype{VIII},
Ne\emissiontype{X}, and Mg\emissiontype{XI} lines showed clear
enhancements. There are also features in the 750 to 900~eV range that
could be due to some combination of Fe\emissiontype{XVII} and
\emissiontype{XVIII} L-lines, higher order transitions of
O\emissiontype{VIII} ($3p$ to $1s$ and $6p$ to $1s$), and a
Ne\emissiontype{IX} line. From the correlation of the X-ray intensity
with solar-wind flux on time scales of about half a day, and from the
short-term ($\sim 10$~minutes) variations of the X-ray intensity, these
lines most likely arise from solar-wind heavy ions interacting with
neutral material in the Earth's magnetosheath. A hard power-law
component is also necessary to explain the LTE spectrum. The origin of
this component is not yet known. Our results indicate that solar
activity can significantly contaminate {\it Suzaku} cosmic X-ray spectra
below $\sim 1$~keV. Recommendations are provided for recognizing such
contamination in observations of extended sources.
\end{abstract}

\section{Introduction}

\citet{Snowden_etal_1994} reported the existence of mysterious X-ray
contamination episodes in the {\it ROSAT} all sky survey data which they
termed long-term enhancements (LTEs).  During an LTE, the X-ray counting
rate in the lower energy bands as much as doubled on a time scale of
1--2~days.  However, they could not find any correlation with other
observational parameters, such as spacecraft position or look direction.
New insight on LTEs was obtained by the discovery of X-ray emission from
comet Hyakutake (\cite{Lisse_etal_1996}).  Following the discovery,
X-rays were detected from many comets (e.g.,
\cite{Dennerl_etal_1997,Cravens_2002}), and the emission mechanism is
now well understood as charge exchange of solar-wind heavy ions with
cometary neutrals (see \cite{Krasnopolsky_etal_2004} for a review).
Then \citet{Cox_1998} and \citet{Cravens_2000} suggested that solar-wind
charge exchange with neutrals in the Earth's geocorona and in the
heliosphere accounts for a part of soft X-ray background below 1~keV.
\citet{Robertson_etal_2001} showed that the LTEs of the {\it ROSAT} all
sky survey data were well correlated with the solar-wind proton flux,
which strongly suggests the origin of the LTEs to be solar-wind charge
exchange with H\emissiontype{I} in the geocorona.  Solar-wind charge
exchange results in line emission from highly ionized ions
(\cite{Krasnopolsky_etal_2004} and references therein).  {\it ROSAT} did
not have enough spectral resolution to resolve those lines.  Spectral
information on geocoronal solar-wind charge exchange was first obtained
during a {\it Chandra} dark moon observation \citep{Wargelin_etal_2004}.
The X-ray photons detected in the direction of the dark moon are most
likely from this source.  The emission spectrum could be described by a
sum of C\emissiontype{VI}, O\emissiontype{VII}, and O\emissiontype{VIII}
K-lines, although the statistics and energy resolution were limited.
More recently, \citet{Snowden_etal_2004} reported time variation of soft
X-ray intensity during the {\it XMM-Newton} {\it Hubble} deep field
north observation. The enhancement was correlated with solar-wind proton
flux variations. They detected C\emissiontype{VI}, O\emissiontype{VII},
O\emissiontype{VIII}, Ne\emissiontype{IX}, and Mg\emissiontype{XI}
emission lines in the enhancement.

The importance of solar-wind charge-exchange emission is three-fold.
Firstly, it will enable us to remotely study low density neutrals in
geocorona, in outer atmosphere of planets such as Jupiter, in
interplanetary space, and in particular, around comets.  Secondly, it
can be used as a highly sensitive ion probe for the solar wind.  Perhaps
most importantly, it becomes a significant contaminating
foreground in the study of the cosmic soft X-ray background below
$\sim$ 1~keV.  More than half of the soft X-ray background at these
energies is considered to arise from hot gas in the disk and halo of our
galaxy and in intergalactic space.  Although such emission was detected
early in the history of X-ray astronomy (e.g. \cite{Tanaka_1977}), its
origins and the physical state of the hot gas are not yet well
understood.  Geocoronal interactions with the solar wind produce at
least a sporadic contamination that must be avoided in observations of
any extended source.  Charge exchange with interstellar neutrals moving
through the heliosphere creates a more subtle contribution where much of
the diagnostic rapid time variation is washed out by the large travel
time of solar-wind events through interplanetary space, and the spectral
lines arise from the same ions expected in hot interstellar plasmas.
\citet{Lallement_2004} used a simplified model of interplanetary charge
exchange to estimate that essentially all of the minimum flux observed
near the Galactic plane in the 1/4~keV (R12) band of the {\it ROSAT} sky
survey could arise from this source (see also
\cite{Pepino_2004}).  We are far from an understanding of solar-wind
charge exchange adequate to determine the true extent of this
contribution.

The X-ray Imaging Spectrometer (XIS) on board {\it Suzaku}
\citep{Suzaku} has a significantly improved spectral line response
function compared to previous X-ray CCD cameras, particularly below
1~keV \citep{XIS}. Together with the large effective area of the X-ray
telescopes (XRT; \cite{XRT}), this will open a new era for the study of
soft X-ray background.  In this paper, we report on {\it Suzaku}
observations of a blank field in direction of the north ecliptic pole
(NEP).  We detected a significant enhancement of the soft X-ray flux
lasting for $\sim 10$~hours.  The enhancement is mostly explained by
increases in C\emissiontype{VI} through Mg\emissiontype{XI} emission
lines.  During the enhancement, both C\emissiontype{VI} $n=2$ to 1 and
C\emissiontype{VI} $n=4$ to 1 transition lines were clearly detected,
which is a firm evidence for charge-exchange emission.  We consider that
the emission is due to the charge-exchange interaction of solar-wind
heavy ions with neutrals in the magnetosheath at 2--8 Earth radii
($R_{\oplus}$).  In this paper, we will concentrate on the X-ray spectra
and the emission processes, and their implications for cosmic X-ray
observations.  The geophysical implications of the results will be
discussed in a separate paper.

Errors quoted in the text and tables are at 90\% confidence single
parameter errors and at $1\sigma$ confidence level in the figures,
unless otherwise stated.

\section{Analysis and results}

The NEP region was observed with {\it Suzaku} twice during the Science
Working Group (SWG) observation time. The XIS was set to normal clocking
mode and the data format was either $3\times 3$ or $5\times 5$. A log of
the observations is shown in table~\ref{tab:obslog}. In this paper, we
concentrate on the spectral change during a ``flare'' detected in the first
observation. For that purpose, we try to model the stable components,
and then evaluate the spectral change.  We use the data from the
backside-illuminated CCD (XIS1), because of its much superior
performance below 1~keV \citep{XIS}.

\begin{table*}
\caption{Log of the NEP observations}
\label{tab:obslog}
\begin{center}
\begin{tabular}{c|cc}\hline\hline
target coordinates & \multicolumn{2}{c}{($\alpha$, $\delta$) =
 (272.8000, 66.0000)}\\
observation ID & 100018010 & 500026010\\
observation period & 2005 Sep. 2 14:30--Sep. 4 15:00 & 2006
	 Feb. 10 5:50--Feb. 12. 2:00\\
net exposure time & 109.8~ks & 88.6~ks\\
 & ($3\times 3$: 95.7~ks, $5\times 5$: 14.1~ks)  &
 ($3\times 3$: 71.5~ks, $5\times 5$: 17.1~ks) \\
\hline
\end{tabular}
\end{center}
\end{table*}

\subsection{Data reduction}
\label{sec:data_reduction}

We used version 0.6 processed {\it Suzaku} data\footnote{Version 0
processing is an internal processing applied to the Suzaku data obtained
during the SWG phase, for the purpose of establishing the detector
calibration as quickly as possible. Some processes that are not critical
for most of the initial calibration and scientific studies, e.g., aspect
correction, fine tuning of the event time tagging of the XIS data, are
skipped in version 0 processing, and hence, quality of the products is
limited in these directions, compared with the official data supplied to
the guest observers. As of 2006 July, version 0.7 is the latest, where
the absolute energy scale accuracy of $\pm0.2$~eV at the
iron K$\alpha$ energy and $\pm5$~eV below 1~keV is achieved for the XIS
data \citep{XIS}. In this paper, we used version 0.6 data where the
energy scale of the XIS data are less accurate ($\sim10$~eV
below 1~keV) than that of version 0.7, because the
empirical model of the contamination distribution was obtained based on
version 0.6 data. Instead, we adjusted the scale and the offset of the
response matrix, as shown in section~\ref{sec:data_reduction}.}.  In
addition to the standard data selection criteria: elevation from sunlit
earth rim $> 20^{\circ}$, elevation from dark earth rim $> 5^{\circ}$,
we applied cutoff rigidity (COR) $>8$ to clean the XIS data.  The XIS
pulse height data for each X-ray event ($3\times 3$ or $5\times 5$
format) were converted to PI (Pulse Invariant) channels using the
`xispi' software version 2005-12-26 and CTI parameters from 2006-01-25.

\begin{figure}
  \begin{center}
\FigureFile(80mm,120mm){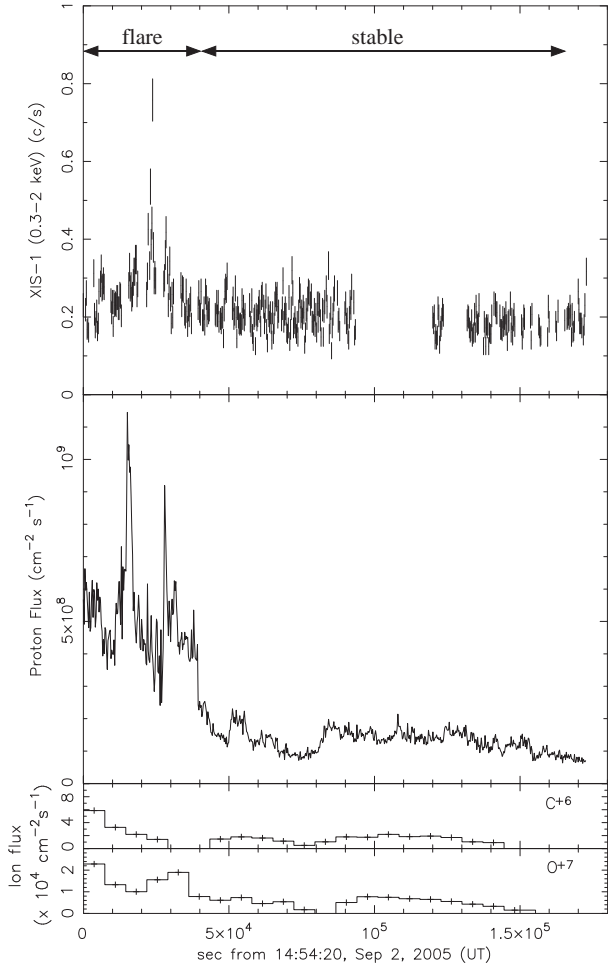}
  \end{center}
\caption{XIS1 counting rate in the 0.3--2 keV energy band (top panel),
solar-wind proton flux (second), C$^{+6}$ flux (third), and O$^{+7}$
flux (bottom) as a function of time.  The {\it Suzaku} XIS counting rate
is shown in 256 s bins.  Particle background counts are not
subtracted. The solar-wind proton flux was calculated using level 2 ACE
SWEPAM data. Each bin of the ACE data was first shifted in time to
correct for the travel time of the solar wind from ACE to the Earth
(typically $\sim 2700$~s), then rebinned into 256 s bins. The ion fluxes
(C$^{+6}$, O$^{+7}$) were calculated from level 2 ACE SWICS data. Only
good data with quality flag 0 were used. See also text in
section~\ref{sec:discussion}.}  \label{fig:light_curve}
\end{figure}

We first created a time history of the X-ray counting rate by binning
the event data into 256~s time bins.  In figure~\ref{fig:light_curve},
we show the 0.3--2~keV counting rate of XIS1 where the non-X-ray
(particle-induced) background rate is not subtracted.  The counting rate
shows a clear enhancement in the first $\sim 4 \times 10^4$~s.  The
particle background rate of the XIS is known to vary because the cosmic
ray flux changes as a function of the spacecraft position over the
Earth.  The background rate is well reproduced as a consistent function
of the local cutoff rigidity.  From the XIS data during intervals when
the telescope is pointed at the dark side of the Earth, we found that
the 0.3--2~keV XIS1 non-X-ray counting rate varies only from
0.03~cts\,s$^{-1}$ to 0.07~cts\,s$^{-1}$ when the cutoff rigidity varies
from 15 to 6~GV.  Thus the observed enhancement cannot
be particle background variation.

\begin{figure}
  \begin{center}
    \FigureFile(80mm,80mm){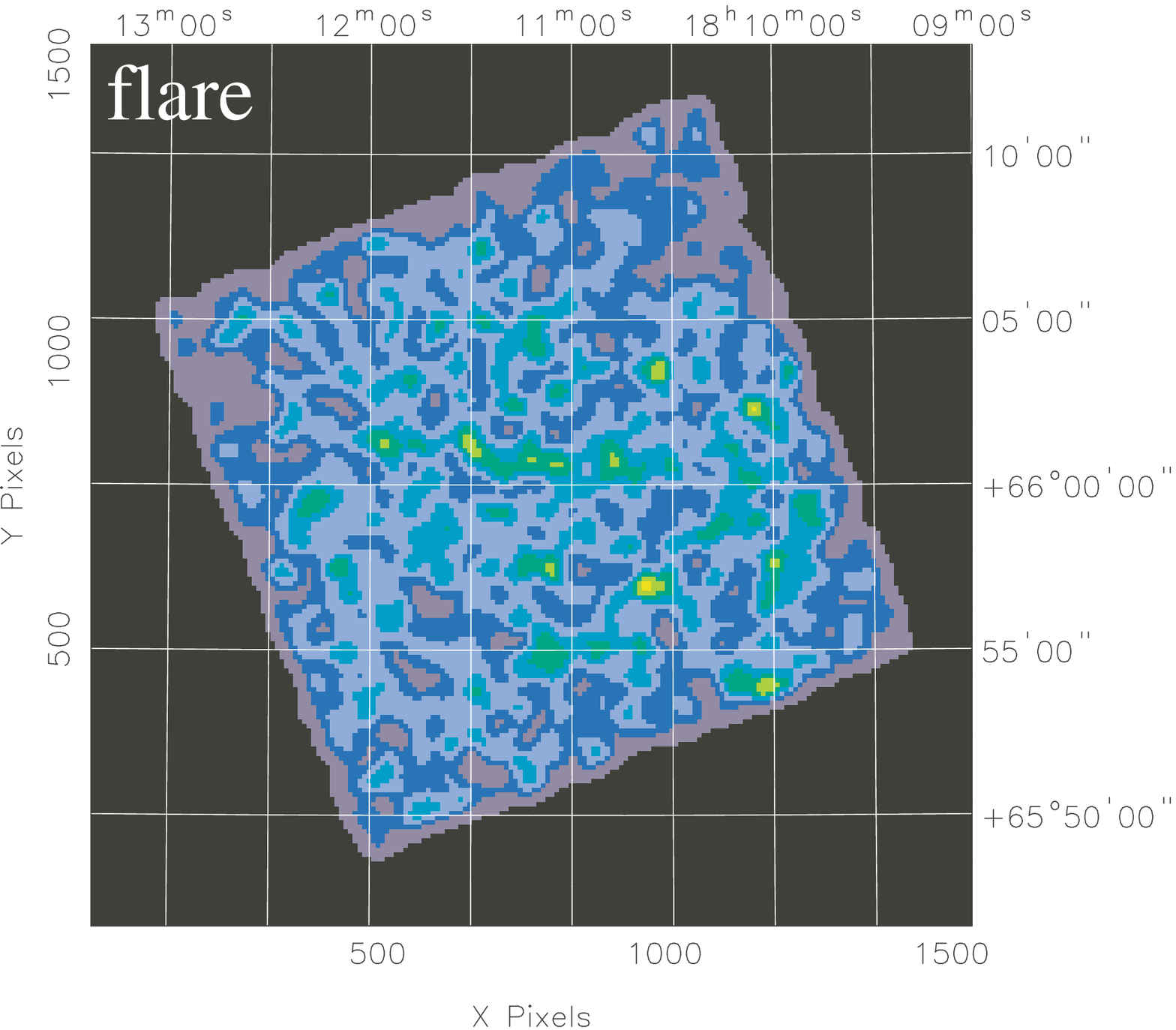}
    \FigureFile(80mm,80mm){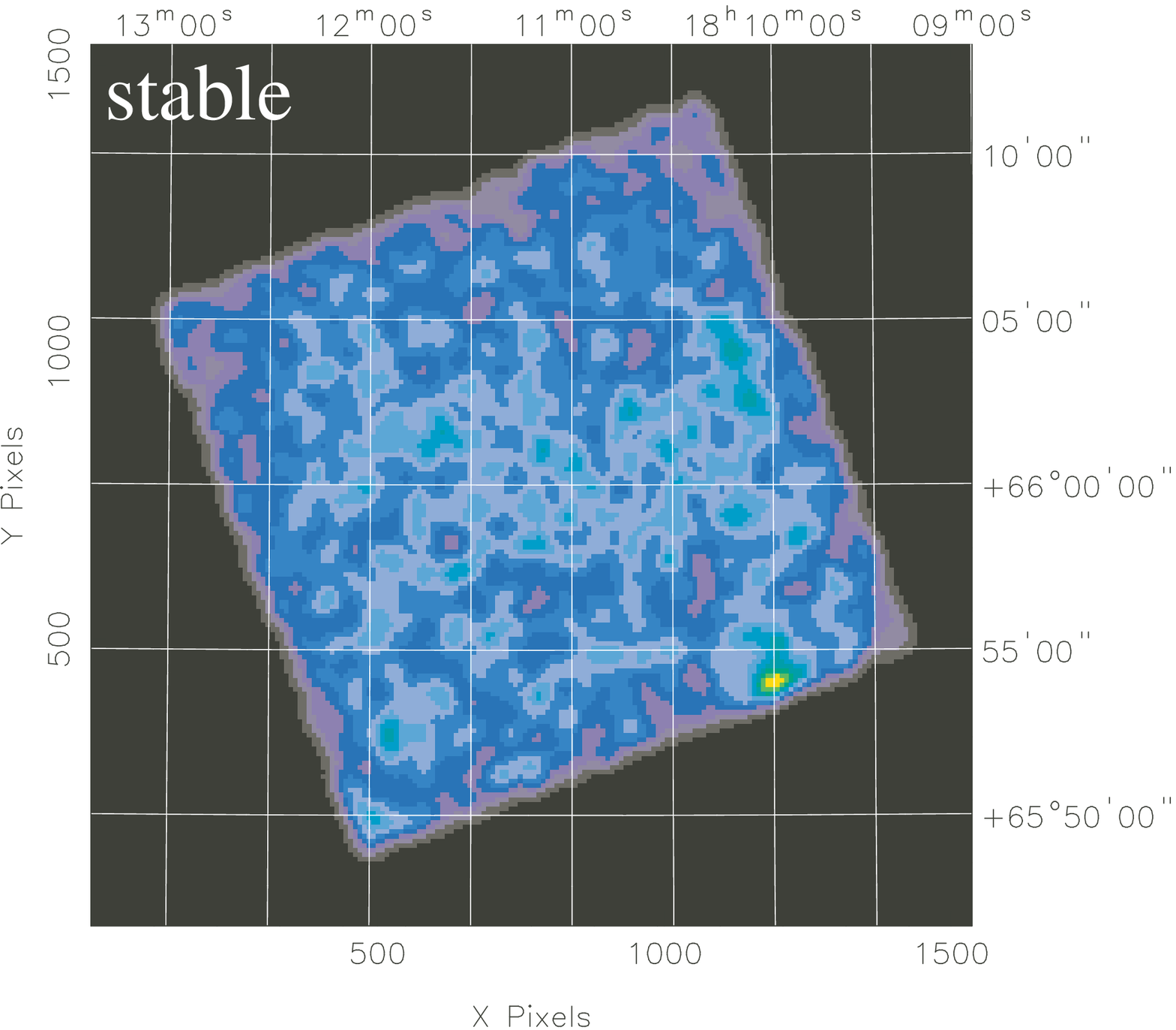}
  \end{center}
\caption{XIS1 images in the 0.3--2~keV band for ``flare'' and ``stable''
periods.}
\label{fig:images}
\end{figure}

The enhancement lasted for $\sim 10$~hours.  Within the enhancement
there are shorter time variations. For example, there are sharp spikes
just before and after the highest peak at $\sim 2.2 \times 10^4$~s.  A
time scale of these variations is as short as $\sim 10$~minutes. We
defined the ``flare'' interval to be 0 to $4\times 10^4$~s and the
``stable'' interval as $4\times 10^4$ to $16.5\times 10^4$~s as shown in
figure~\ref{fig:light_curve}, then created X-ray images for ``flare''
and ``stable'' periods separately. The images in figure~\ref{fig:images}
show that the enhancement is not due to changes in any discrete sources.

We subtracted the particle background from both ``stable'' and ``flare''
spectra using the dark earth data with the same distribution of cutoff
rigidities.  The counting rates of the background subtracted spectra
above 10~keV, where the data are dominated by non X-ray background
events, are $(2.9\pm 0.5)\times 10^{-2}$~cts\,s$^{-1}$ for ``flare'' and
$(1.2\pm 0.3)\times 10^{-2}$~cts\,s$^{-1}$ for ``stable'', where the
error is $1\sigma$ statistical error.  These correspond to 6\% and 2\%
of the counting rates of the dark earth data, respectively. Note that,
in the 2--5~keV band, change of the spectral model due to this
background difference is much smaller than the statistical error, and
hence, they do not affect the results in the later sections.

Since we suppose the diffuse X-ray emission to be approximately uniform
over the field of view of the XIS, we created an X-ray telescope (XRT)
response function, or more specifically, an ancillary response file
(ARF) used in the XSPEC spectral fit software \citep{xspec}, for a
flat field, using the XIS ARF builder software {\it xissimarfgen}
\citep{xissimarfgen}.  It is known that contamination has been
accumulating on the optical blocking filters of the XIS sensors since
the detector doors were opened following the launch, and that it
accumulates much more quickly at the center of the field of view than at
the outside \citep{XIS}. We used the contaminant thickness and radial
distribution functions version 2006-05-28 when we built an ARF with
{\it xissimarfgen}. At the early time of this NEP observation, the contaminant
column densities were only $4.1\times 10^{17}$~carbon~atoms~cm$^{-2}$
and $0.7\times 10^{17}$~oxygen~atoms~cm$^{-2}$ at the center for XIS1.
This reduces the efficiency by about 12\% at the energy of
C\emissiontype{VI} (367~eV), 6\% at O\emissiontype{VII} (561~eV), 4\%
at O\emissiontype{VIII} (653~eV), and $<2$\% at Ne\emissiontype{IX}
(905~eV) and higher energies. Systematic errors in the contaminant
thickness are estimated to be about 10\%.  The transmission uncertainty
due to this systematic error is only 1\% for C\emissiontype{VI}, and
less for lines at higher energies; hence it is negligible compared with
other errors.  For the XIS response function, we used the
ae\_xi1\_20060213c.rmf file supplied by the XIS team, with energy scale
corrections of slope 0.9948 and offset $-0.0035$~keV as determined
through the iterative analysis described in the next section.

\subsection{Spectral fit of ``stable'' spectrum}

We then performed spectral fits to the ``stable'' spectrum. We first
restricted the fitting energy range to 2--5~keV.  In this range the
emission is dominated by the Cosmic X-ray Background (CXB), which is
largely emission from unresolved AGNs and can be represented by a power
law function absorbed by neutral material along the line of sight
through our Galaxy. We thus fitted the spectrum with a power-law
function with absorption by a neutral medium with solar abundances
\citep{AG89} and fixed the absorbing column density at the total
Galactic value in this direction $N_{\rm H} = 4.4 \times 10^{20}~{\rm
cm}^{-2}$ \citep{NH}\footnote{We used `nH' tool available at
http://heasarc.gsfc.nasa.gov/cgi-bin/Tools/w3nh/w3nh.pl.}.  The fit
results are summarized in the second column of
table~\ref{tab:fits_stable}.  The photon index is consistent with the
nominal CXB value ($1.40\pm 0.05$; \cite{Marshall_etal_1980}).  The
normalization of the power-law function is also consistent with previous
observations; 9--11~photons\,cm$^{-2}$\,s$^{-1}$\,sr$^{-1}$\,keV$^{-1}$
at 1~keV (e.g., \cite{Gendreau_etal_1995,Revnivtsev_etal_2005}).

\begin{table*}
\caption{Results of spectral fits to the ``stable'' spectrum}
\label{tab:fits_stable}
\begin{center}
\begin{tabular}{llll}\hline\hline 
\hspace{5mm} & Parameter  & 2--5~keV & 0.2--2~keV \\\hline
\multicolumn{4}{l}{Power-law component}\\
\hline
 & $N_{\rm H}$ [$10^{22}$~cm$^{-2}$] & 0.044 (fixed) & 0.044 (fixed)\\
     & photon index $\Gamma$ & $1.33\pm 0.23$  & 1.33 (fixed) \\
 &normalization\footnotemark[$*$] &  $10.4^{+3.1}_{-2.4}$ &
     10.4 (fixed) \\
\hline
\multicolumn{4}{l}{Thin-thermal component (VMEKAL)}\\
\hline
 &  $kT$ [keV] &         ---   & $0.177^{+0.003}_{-0.002}$\\
 & C abundance (solar) & ---   & $1.92^{+0.40}_{-0.36}$\\
 & N abundance  (solar)& ---   & $2.14^{+0.33}_{-0.31}$\\
 & O abundance  (solar)&  ---  & 1.0 (fixed)\\
 & Ne abundance  (solar)& ---  & $2.77^{+0.53}_{-0.59}$\\
 & Fe abundance  (solar)& ---  & $1.42^{+0.20}_{-0.22}$\\ 
  &  normalization\footnotemark[$\dagger$] & --- & $16.35^{+0.62}_{-0.68}$\\
\hline
\multicolumn{2}{l}{gain slope} & 0.9948 (fixed)   & 0.9948 \\
\multicolumn{2}{l}{gain offset}   & $-0.0035$ (fixed) & $-0.0035$\\
\hline
\multicolumn{2}{l}{$\chi^2$/degrees of freedom} & 38.39/38 & 280.15/228\\
\hline
\multicolumn{3}{@{}l@{}}{\hbox to 0pt{\parbox{95mm}{\footnotesize
\footnotemark[$*$] In units of photons\,cm$^{-2}$\,s$^{-1}$\,sr$^{-1}$\,keV$^{-1}$ at 1~keV.
\par\noindent
\footnotemark[$\dagger$] $(4\pi)^{-1}D_{\rm A}^{-2}(1+z)^{-2} 10^{-14}\int n_{\rm e}
 n_{\rm H}dV$ per steradian, where $D_{\rm A}$ is the angular size
 distance to the source (cm), and $n_{\rm e}$, $n_{\rm H}$ are the
 electron and hydrogen densities (cm$^{-3}$), respectively.
}\hss}}
\end{tabular}
\end{center}
\end{table*}

\begin{figure}
  \begin{center}
  \FigureFile(80mm,60mm){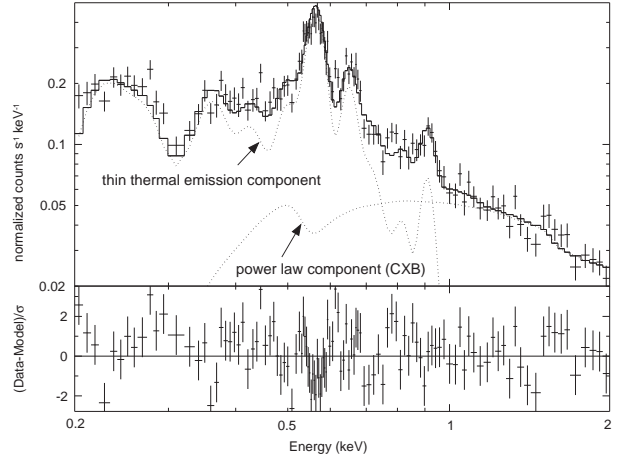}
  \end{center}
\caption{Spectral fit to the ``stable'' spectrum in the 0.2--2~keV range.
Observed spectrum is plotted in the upper panel with crosses where the
vertical bars correspond to $1\sigma$ statistical errors. The thick
step function is the best-fit model function convolved with the X-ray
mirror and the detector response functions. The dotted lines show
contributions of different spectral components. The lower panel shows the
residuals of the fits.
}\label{fig:spec_stable}
\end{figure}

We then fitted the spectrum over the 0.2--2~keV range.  As shown in
figure~\ref{fig:spec_stable}, it is clear that there is additional emission
below 1 keV and the excess contains emission lines such as
O\emissiontype{VII}, O\emissiontype{VIII} and Ne\emissiontype{IX}
K$\alpha$.  Thus, fixing the spectral parameters of the CXB component to
the best fit values of the 2--5~keV fit, we added a thin thermal
emission component using the MEKAL model
(\cite{Mewe_1985,Mewe_1986,Kaastra_1992,Liedahl_1995}).  We fixed the
abundance of O at the solar value, and set abundances of other elements
(C, N, Ne, Fe) free. The version 0.6 XIS data products are known to
contain systematic energy calibration errors of $\sim 10$~eV amplitude
below $\sim 2$~keV.  We therefore adjusted the energy scale by varying
the gain slope and offset as free parameters of the fit. We determined
the gain and the offset in this fitting, and applied the same gain and
offset throughout the paper, including the fitting of the CXB component
described in the previous paragraph. The results are shown in the third
column of table~\ref{tab:fits_stable}.  The model adequately represent
the observed spectrum\footnote{There are small positive residuals,
especially at around 300~eV and 450~eV. We can model these features with
additional delta functions of variable energy and amplitude. Even if we
add them, however, normalizations of the delta functions employed for
the ``flare'' spectrum (section~\ref{sec:flare spectrum}) are affected
by 10\% at most.}. Therefore, we adopt the model shown in
table~\ref{tab:fits_stable} as a representative model for the ``stable''
spectrum, in order to evaluate the spectral change during the ``flare''.

The spectrum could be fit by other thermal models.  When we adopt the
present single temperature model with varying abundances, the
temperature is determined primarily by the O\emissiontype{VII} to
O\emissiontype{VIII} K$\alpha$ emission line intensity ratio, and the
abundances are determined by the intensities of C\emissiontype{VI}
K$\alpha$, N\emissiontype{VI} K$\alpha$, Fe\emissiontype{XVII}-L and
Ne\emissiontype{IX} K$\alpha$.  If we employ a multi-temperature thermal
model, it may not be necessary to vary abundances because line intensity
ratios can be adjusted by choosing temperatures.  The most important
result here is that the excess emission above the CXB below 1~keV can be
represented by the emission lines of a thermal model.  We tried to
include additional continuum emission represented by a power-law
function or thermal bremsstrahlung model. However, there was no
improvement in $\chi^2$.

\subsection{Spectral fit of ``flare'' spectrum}
\label{sec:flare spectrum}

\begin{table*}
\caption{Results of spectral fits to the ``flare'' spectrum. Parameter
values of components added to the ``stable'' spectral model.}
\label{tab:fits_flare}
\begin{center}
\begin{tabular}{lllll}\hline\hline
\hspace{5mm} & Parameter  & 2--5~keV & 0.2--2~keV & 
Line identification\footnotemark[$*$] \\\hline
\multicolumn{5}{l}{Additional Power-law component without absorption}\\
\hline
 & photon index $\Gamma$ & $0.0\pm 1.7$ & 0.0 (fixed) \\
 & normalization\footnotemark[$\dagger$]   &  $7.4^{+8.8}_{-1.3}\times 10^{-1}$ & \multicolumn{2}{l}{0.74 (fixed) } \\
\hline
\multicolumn{5}{l}{Additional narrow Gaussian lines}\\
\hline
1 &  center energy [eV] & --- & $269\pm 4$ & C band lines \\
  & normalization\footnotemark[$\ddagger$] & --- & $7.7^{+1.8}_{-1.7}$& \\
2 & center energy [eV] & --- & $357^{+6}_{-8}$ & C\emissiontype{VI}  2p to 1s (367 eV)\\
    & normalization\footnotemark[$\ddagger$] & --- & $7.3^{+2.2}_{-1.4}$& \\
3 & center energy [eV] & --- & $455^{+5}_{-13}$ & C\emissiontype{VI}  4p to 1s (459 eV) \\
   & normalization\footnotemark[$\ddagger$] & --- & $3.09^{+0.74}_{-0.76}$ \\
4 & center energy [eV] & --- & $558^{+8}_{-9}$ &  O\emissiontype{VII}
		 (561~eV) \\
   & normalization\footnotemark[$\ddagger$] & --- &$5.1^{+1.1}_{-1.0}$ & \\
5 & center energy [eV] & --- & $649^{+4}_{-6}$ & O\emissiontype{VIII} 2p to 1s (653 eV) \\
   & normalization\footnotemark[$\ddagger$] & --- &  $5.02^{+0.58}_{-0.76}$ \\
6  & center energy [eV] & --- & $796^{+10}_{-8}$ &  Fe\emissiontype{XVII,XVIII}-L + O\emissiontype{VIII} 3p to 1s (774 eV)?\\
   & normalization\footnotemark[$\ddagger$] & --- &$1.67^{+0.35}_{-0.34} $ &\\
7 & center energy [eV] & --- &$882^{+14}_{-17}$ &Fe\emissiontype{XVII,XVIII}-L +
		 Ne\emissiontype{IX} (905~eV) \\
   & normalization\footnotemark[$\ddagger$] & --- &$0.95^{+0.26}_{-0.33}$ & + O\emissiontype{VIII} 6p to 1s (847 eV)?\\
8  & center energy [eV] & --- &$1022^{+11}_{-7}$  &Ne\emissiontype{X} (1022~eV) \\
   & normalization\footnotemark[$\ddagger$] & --- &$1.04^{+0.20}_{-0.29}$ \\
9 & center energy [eV] & --- &$1356^{+16}_{-20}$ & Mg\emissiontype{XI}
		 (1329~eV)  \\
   & normalization\footnotemark[$\ddagger$] & --- & $0.73^{+0.19}_{-0.20}$ \\
\hline
 \multicolumn{2}{l}{$\chi^2$ /d.o.f} & 17.65/14 & 161.85/114\\
\hline
\multicolumn{4}{@{}l@{}}{\hbox to 0pt{\parbox{160mm}{\footnotesize
\footnotemark[$*$] Line energies at the rest frame are taken from
\citet{Kharchenko_2003}, \citet{Krasnopolsky_2004}.  Energies of the
forbidden line are shown for O\emissiontype{VII}, Ne\emissiontype{IX},
and Mg\emissiontype{XI} K$\alpha$, because the forbidden line becomes much
stronger at the charge-exchange emission (e.g., \cite{Kharchenko_2003}).
\par\noindent
\footnotemark[$\dagger$] In units of photons\,cm$^{-2}$\,s$^{-1}$\,sr$^{-1}$\,keV$^{-1}$ at
 1~keV.
\par\noindent
\footnotemark[$\ddagger$] In units of photons\,cm$^{-2}$\,s$^{-1}$\,sr$^{-1}$.
}\hss}}
\end{tabular}
\end{center}
\end{table*}

\begin{figure}
  \begin{center}
   \FigureFile(80mm,60mm){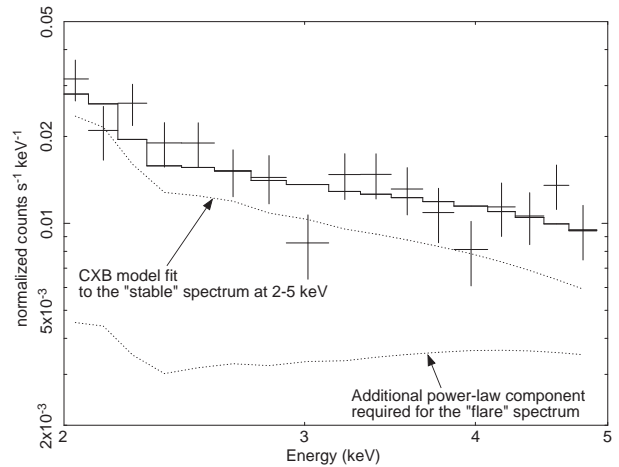}
  \end{center}
\caption{Comparison of the ``flare'' spectrum with the CXB model fit to
the ``stable'' spectrum at 2--5~keV.  An additional power-law component
was required in this range for the ``flare'' spectrum.  Parameters are
given in the second column of table~3.}
\label{fig:spec_flare_2-5}
\end{figure}

\begin{figure}
  \begin{center}
   \FigureFile(80mm,60mm){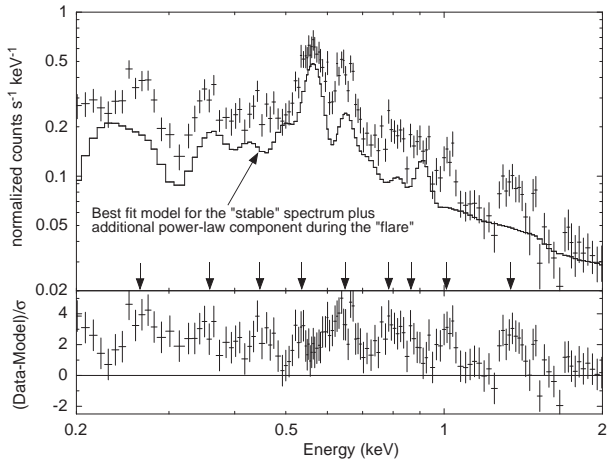}
  \end{center}
\caption{Comparison of the ``flare'' spectrum with the best fit model
for the ``stable'' spectrum plus additional power-law component during
the ``flare'' in the 0.2--2~keV energy range. Vertical arrows
indicate line-like structures in the residuals. }
  \label{fig:spec_compare}
\end{figure}

\begin{figure}
  \begin{center}
   \FigureFile(80mm,60mm){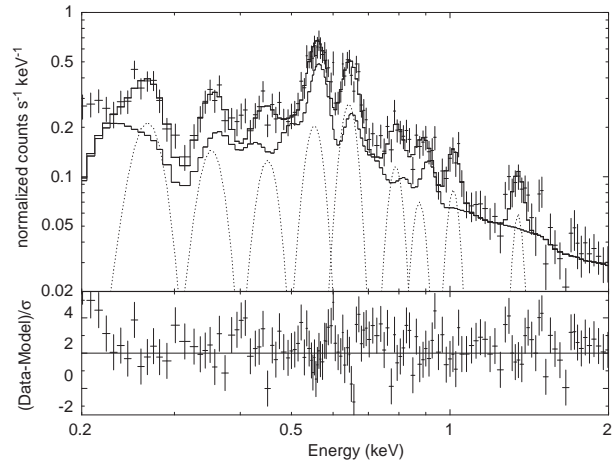}
  \end{center}
\caption{``Flare'' spectrum and best-fit model spectrum.  The model
spectrum is a sum of the best fit model for the ``stable'' spectrum plus
additional power-law component during the ``flare'' shown with a thin
solid curve (same as that shown in figure~\ref{fig:spec_compare}), and
nine emission lines shown with dotted curves. Best-fit parameters of the
nine emission lines are summarized in table~\ref{tab:fits_flare}.}
\label{fig:spec_flare}
\end{figure}

We first compared the ``flare'' spectrum and the best fit model for the
``stable'' spectrum in the 2--5~keV range. We found that there is an
excess hard emission ($\chi^2$/d.o.f $=46.21/16$) and added an
additional power law component, as shown in figure~4.  Then the ``flare''
0.2--2~keV spectrum was compared with a model consisting of the best fit
``stable'' model and the additional power law component as shown in
figure~\ref{fig:spec_compare}.  The residuals at the bottom of the
figure show line-like structures, so we have added nine emission lines
as indicated by the arrows.  All lines were modeled by delta functions
of variable energy and amplitude. We show the results in the third
column of table~\ref{tab:fits_flare} and
figure~\ref{fig:spec_flare}. The ``flare'' spectrum is well represented by
the model\footnote{The least significant line is that at 882~eV. By
adding this line, $\chi^2$/d.o.f was improved from 188.61/116 to
161.85/114. This line is significant at a greater than 99.98\%
confidence level based on the $F$-test.}. Therefore, the enhancement of
the X-ray intensity during the ``flare'' can be explained by an increase
in these emission lines and the hard power-law emission.

In the fourth column of table~\ref{tab:fits_flare}, we show probable
identifications of lines.  The lowest energy line below the carbon edge
at $269\pm4$~eV is considered to be a sum of multiple L emission
lines. The line at $455^{+5}_{-13}$~eV is most likely the $n=4$ to 1
transition (Ly$\gamma$) of C\emissiontype{VI} at 459~eV, since the line
energy is consistent within the statistical error and there are no other
likely emission lines at this energy. If we introduced two lines at
436~eV (C\emissiontype{VI} Ly$\beta$; $n=3$ to 1) and 459~eV
(C\emissiontype{VI} Ly$\gamma$) instead of one line of variable energy,
the normalizations of these lines became $1.93^{+0.75}_{-0.99}$ and
$1.68^{+0.76}_{-0.71}$~photons\,cm$^{-2}$\,s$^{-1}$\,sr$^{-1}$,
respectively. Therefore, C\emissiontype{VI} Ly$\beta$ could have a
comparable contribution. In either case, we definitely need the
C\emissiontype{VI} Ly$\gamma$ line. \citet{Dennerl_2003} attributed a
weak peak structure found in the XMM-Newton spectrum of comet C/2000 to
a sum of C\emissiontype{VI} Ly$\beta$ and Ly$\gamma$, and
C\emissiontype{VI} Ly$\gamma$ due to charge-exchange emission between
the highly ionized solar wind and exospheric or interplanetary neutrals
during an XMM-Newton observation of the Hubble Deep Field--North was
reported by \citet{Snowden_etal_2004}.  This NEP observation, however,
seems to be the clearest detection so far.  We also have detected
O\emissiontype{VII} to Mg\emissiontype{XI} lines. The lines at 796 and
882 eV~are likely to represent complex structures due to Fe-L and other
lines.

\section{Discussion}
\label{sec:discussion}

The short ($\sim 10$~minutes) time scale variations observed during the
enhancement of X-ray intensity imply that the size of
the emission region is no larger than 10 light minutes. On the other
hand, the apparent size of the emission region must be equal to or
larger than the XIS field of view ($18'$). These require the emitter of
the X-ray enhancement to be within a distance of $10~{\rm
light~minutes}/18'$, or $\sim 10^{-3}$~pc.  Because the enhanced X-ray
emission consists of emission lines from C\emissiontype{VI} to
Mg\emissiontype{XI}, this requires an ion source within
$10^{-3}$~pc. There is only one ion source in this distance range. That
is the Sun.
 
The Sun may produce X-ray emission lines in our observations in two
possible ways: scattering of solar X-rays by the Earth's atmosphere, and
solar-wind charge exchange.  In the former case, the X-ray intensity is
proportional to the solar X-ray intensity multiplied by the sunlit
atmospheric column density.  The solar X-ray intensity is continuously
monitored by the GOES (Geostationary Operational Environmental
Satellites)\footnote{The data available at
http://www.ngdc.noaa.gov/stp/GOES/goes.html.}, but data show no
correlation with the enhancement.  Moreover, using the MSIS atmosphere
model \citep{Hedin_1991}, we found that the column density of sunlit
atmosphere varied by many orders of magnitude ($10^{9}$) during the
observations, but no correlation was found with the observed X-ray
intensity.  Thus scattering of solar X-rays can be excluded.

\begin{figure}
  \begin{center}
   \FigureFile(80mm,60mm){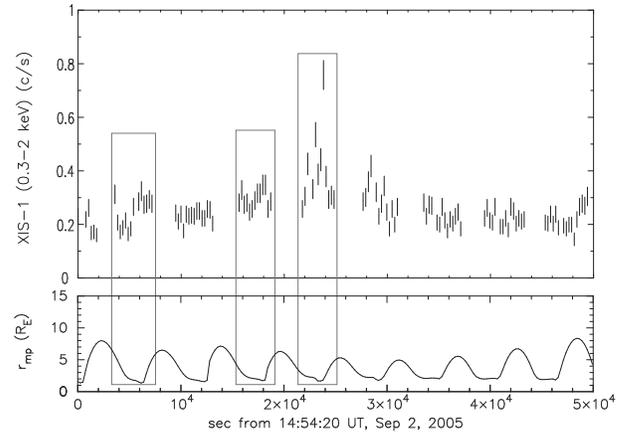}
  \end{center}
\caption{0.3--2 keV X-ray intensity (upper panel) and geocentric
distance of the point whose geomagnetic field becomes open to the space
for the first time along the line of sight from the spacecraft position
in units of the earth radius, as a function of time during the
observation (lower panel). See also the schematic view shown in
figure~\ref{fig:magneto_schematic}.}
  \label{fig:magneto}
\end{figure}

\begin{figure}
  \begin{center}
   \FigureFile(80mm,60mm){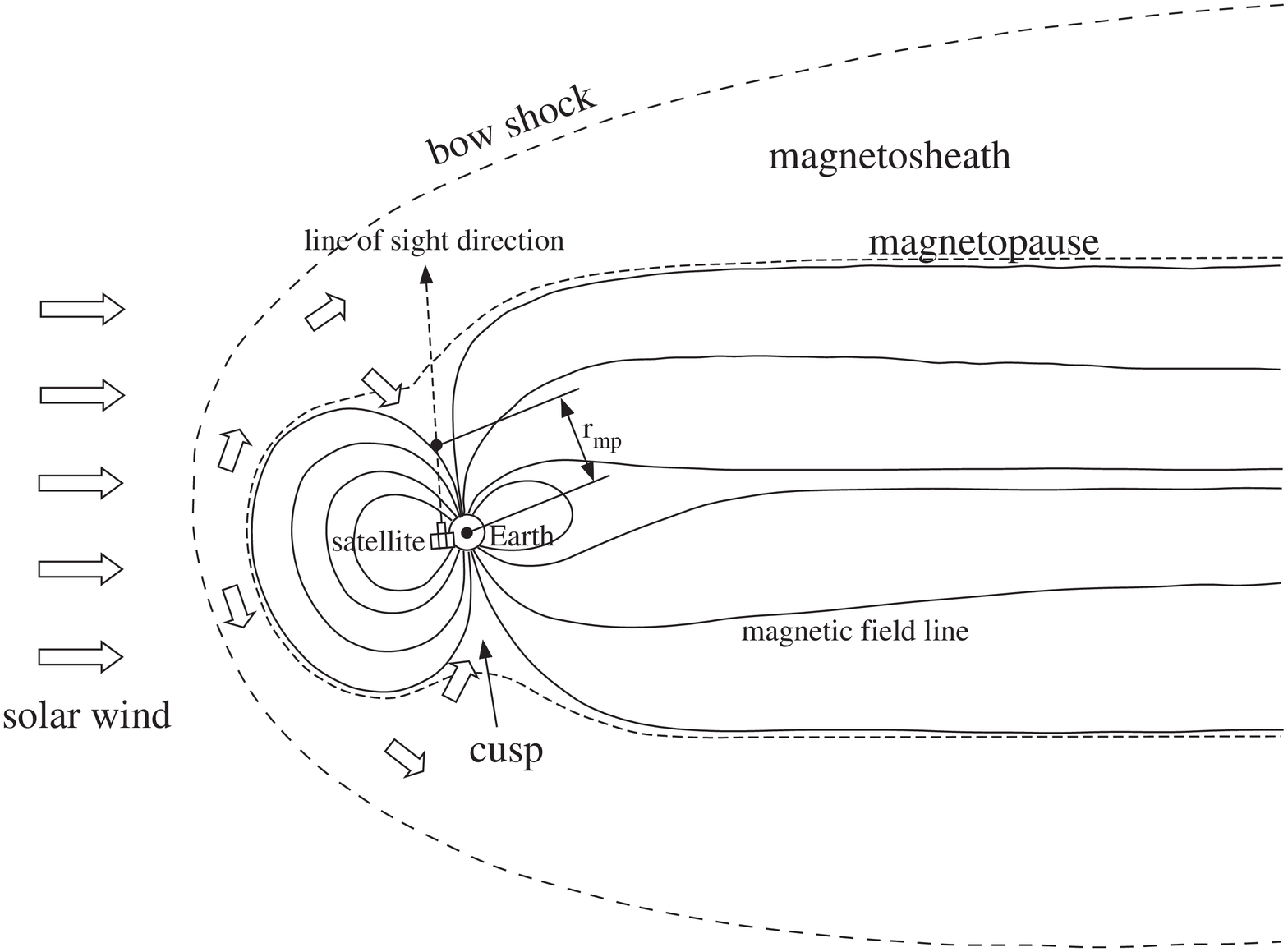}
  \end{center}
\caption{Schematic view of the magnetosphere and the definition of
 $r_{\rm mp}$ used in figure~\ref{fig:magneto}.}
  \label{fig:magneto_schematic}
\end{figure}

In figure \ref{fig:light_curve}, we show the proton flux observed by the
ACE (Advanced Composition Explorer)\footnote{The data available at
http://www.srl.caltech.edu/ACE/ASC/.} together with the {\it Suzaku}
X-ray counting rate.  The ACE data were shifted in time to account for
propagation time from ACE to the Earth.  Clearly the proton flux was
enhanced during the X-ray ``flare''. This is consistent with solar-wind
charge-exchange model.

Charge-exchange X-ray emission also is strongly supported by the
detection of the C\emissiontype{VI} $n=4$ to 1 transition line
(Ly$\gamma$). In CIE (collisional ionization equilibrium) thermal
emission, the Ly$\beta$ and Ly$\gamma$ lines of C\emissiontype{VI} are
suppressed relative to Ly$\alpha$ by the Boltzmann factor in the
distribution of exciting electrons. In charge exchange between
C\emissiontype{VII} and H\emissiontype{I}, the electron is deposited
primarily in the $n=4$ level (\citet{Krasnopolsky_etal_2004} and
references therein), and the X-ray lines are produced when it cascades
to the $n=1$ level, guided only by branching ratios.  In high energy
collisions, angular momentum states are tend to be populated
statistically by weight of the state's degeneracy, and electrons are
primarily captured into maximal $l$, where $n$ can change only by 1 unit
at a time during the cascade \citep{Beiersdorfer_2001}, again resulting
in relatively weak Ly$\beta$ and Ly$\gamma$. Behind the Earth's bow
shock, however, the solar wind velocity is reduced, which should result
in recombination into low $l$-orbitals and strong Ly$\beta$ and
Ly$\gamma$.  In the ``flare'' spectrum, both C\emissiontype{VI}
Ly$\alpha$ and Ly$\gamma$ lines were detected and this is a strong
evidence for charge exchange.  In addition, O\emissiontype{VIII} $n=3$
to 1 and $n=6$ to 1 transition lines are also enhanced in the comet
emission model \citep{Kharchenko_2003}.  However, we cannot separate
those lines from Fe\emissiontype{XVII}-L emission lines with the present
energy resolution.

Although X-ray intensity and the proton flux are correlated on a time
scale of $\sim 10$ hours, they do not show much correlation on short
time scales.  We consider that the short term X-ray intensity variation
is at least partly arising from orbital motion of the spacecraft.  In
figure~\ref{fig:magneto}, we show the geocentric distance of the point
whose geomagnetic field becomes open to the space for the first time
along the line of sight from the spacecraft position, i.e., the point
where the line of sight encounters the magnetosheath (see also the
schematic view shown in figure~\ref{fig:magneto_schematic} for the
definition). We evaluated the end point of the magnetic field using the
software GEOPACK-2005 and T96 magnetic field model
(\citet{Tsyganenko_2005} and references therein)\footnote{The software
package available at
http://modelweb.gsfc.nasa.gov/magnetos/data-based/modeling.html.}.  We
obtained the solar-wind parameters required to perform the calculation
from the CDAWeb (Coordinated Data Analysis Web)\footnote{The data
available at http://cdaweb.gsfc.nasa.gov/cdaweb/sp\_phys/.}.  We find
that the line of sight during the present observation was rather special
in the sense that it goes through the north magnetic pole region where
charged particles of the magnetosheath can penetrate down to
2--$8R_{\oplus}$ moving along open field lines. The short-term X-ray
intensity variations during the time intervals shown by boxes in figure
\ref{fig:magneto} indicate anti-correlation with the distance to the
magnetosheath.  This indicates that the charge exchange of solar-wind
heavy ions is taking place at 2--$8R_{\oplus}$ where the neutral matter
density is high.  \citet{Robertson_etal_2006} recently studied
theoretically solar-wind charge-exchange emission from the
magnetosheath.  Implications of the present results on the solar-wind
ion composition and the Earth's magnetosheath including comparisons with
the theoretical model will be reported in a separate paper.

Finally, since solar-wind charge-exchange emission can become a
difficult foreground in the study of soft diffuse sources, we summarize
a procedure to examine possible contamination in the {\it Suzaku}
spectra below $\sim$ 1~keV by solar activity.
\begin{enumerate}
\item check the light curve, if time variation is not expected for the object,
\item check the solar X-ray intensity and the column density of sunlit
atmosphere along the line of sight,
\item check the solar wind proton flux,
\item check the radius of the magnetosheath on the line of sight.
\end{enumerate}

\bigskip
 
We would like to thank Prof.~K.~Maezawa, and Prof.~I.~Shinohara for
their help in calculations of the Earth's magnetosheath and also for
valuable discussions.  Thanks are also due to Prof.~T.~Mukai,
Prof.~A.~Matsuoka, and Prof.~H.~Hayakawa for discussions on the relation
between {\it Suzaku} data and solar wind. We are grateful to the
referee, Dr.~A.~Dalgarno, for useful comments to improve this paper.

\end{document}